\documentclass[aps,prb,twocolumn,showpacs,groupedaddress,superscriptaddress]{revtex4-1}
\usepackage{amsmath,amssymb}
\usepackage{graphicx}
\usepackage[colorlinks,citecolor=red]{hyperref}
\newcommand\ext{\mathrm{ext}}
\newcommand\eff{\mathrm{eff}}
\newcommand\FQ{\mathrm{FQ}}
\newcommand\FC{\mathrm{FC}}
\newcommand\bfS{\mathrm{S}}
\newcommand\bfR{\mathrm{R}}
\newcommand\mmin{\mathrm{min}} %min is already defined

\newcommand\pk{\mathrm{pk}}

\newcommand\bfH{\boldsymbol{H}}

\newcommand\bfz{\boldsymbol{z}}
\newcommand\bfx{\boldsymbol{x}}

\newcommand\bfmu{\boldsymbol\mu}

\begin{document}

%\preprint{}

\title{Precision determination of a fluxoid quantum's magnetic moment in a superconducting micro-ring}


\author{Heonhwa Choi}
\affiliation{Division of Physical Metrology, Korea Research Institute of Standards and Science, Daejeon 34113, Korea}
\affiliation{Department of Nanoscience, Korea University of Science and Technology, Daejeon 34113, Korea}

\author{Yun Won Kim}
\affiliation{Division of Physical Metrology, Korea Research Institute of Standards and Science, Daejeon 34113, Korea}

\author{Soon-Gul Lee}
\affiliation{Department of Display and Semiconductor Physics, Korea University Sejong Campus, Sejong 30019, Korea}

\author{Mahn-Soo Choi}
\affiliation{Department of Physics, Korea University, Seoul 02841, Korea}

\author{Min-Seok Kim}
\affiliation{Division of Physical Metrology, Korea Research Institute of Standards and Science, Daejeon 34113, Korea}

\author{Jae-Hyuk Choi}
\email[]{jhchoi@kriss.re.kr}
\affiliation{Division of Physical Metrology, Korea Research Institute of Standards and Science, Daejeon 34113, Korea}
\affiliation{Department of Nanoscience, Korea University of Science and Technology, Daejeon 34113, Korea}

\date{\today}

\begin{abstract}
Using  dynamic cantilever magnetometry and experimentally determining the cantilever’s vibrational mode shape, we precisely measured the magnetic moment of a lithographically defined micron-sized superconducting Nb ring, a key element for the previously proposed subpiconewton  force standard. The magnetic moments due to individual magnetic fluxoids and a diamagnetic response were independently determined at $T$ = 4.3 K, with a subfemtoampere-square-meter resolution. The results show good agreement with the theoretical estimation yielded by the Brandt and Clem model within the spring constant determination accuracy. 
\end{abstract}

%\pacs{06.20.fb, 85.25.-j, 84.71.Ba}

\maketitle

\section{INTRODUCTION}

%\subsection{}

% If in two-column mode, this environment will change to single-column
% format so that long equations can be displayed. Use
%\begin{widetext}
%\end{widetext}

%\appendix
%\section{}

%%%%%%%%%%%%%%%%%%%%%%         

The superconducting ring has attracted considerable attention in the context of both fundamental superconductor research and application, because of its geometry-related effects, such as fluxoid quantization and quantum interference.\cite{Tinkham96,Jang11,Chen10,Hasselba08,Kirtley10} 
The magnetic flux, or more precisely, magnetic fluxoid, through an ordinary superconducting ring is quantized in units of $h/2e$, where $h$ is Planck's constant and $e$ is the electron charge.\cite{Tinkham96} In superconducting devices and applications, a superconducting ring with or without Josephson junctions has acted as a key element.\cite{Hasselba08,Kirtley10,Moody02,Weiss15,Choi07} Understanding its magnetic properties is valuable for the design and analysis of, for example, a superconducting quantum interference device (SQUID),\cite{Hasselba08,Kirtley10} a gravity gradiometry,\cite{Moody02} an ultracold atom trap,\cite{Weiss15} and a subpiconewton force standard.\cite{Choi07}  In particular, the concept of quantum-based force realization,\cite{Choi07} which some authors have suggested as a candidate for the subpiconewton force standard previously, utilizes magnetic fluxoid quanta in a microscale superconducting ring. The force can be increased or decreased by a force step, estimated to be on the subpiconewton level, by controlling the fluxoid number. The magnetic moment due to a single fluxoid quantum is the minimum unit for generating a magnetic force in a well-defined magnetic field gradient. 

Determining the unit magnetic moment with not only high sensitivity, but also high precision is key towards establishing the suggested method as the first standard  for an extremely small force, because the unit magnetic moment  defines the magnitude and precision of the unit force to be realized. Besides the small-force-standard application,\cite{Choi07} the unit magnetic moment based on fluxoid quanta can be utilized as a new reference for a small magnetic moment at the femtoampere-square-meter level.

Several theoretical methods\cite{Brandt97,Brojeny03,Brandt04} have been developed to calculate the magnetic moments as well as the magnetic-field and current-density profiles for various values of the fluxoid number and external magnetic field in superconducting thin-film rings and disks. Initially, cases of negligibly small penetration depth $\lambda$ were addressed,\cite{Brandt97,Brojeny03} and Brandt and Clem\cite{Brandt04} generalized the previous studies to finite $\lambda$, providing a calculation method to give precise numerical solutions. Although their theory has been adopted for superconducting ring design or to interpret its properties over the past decade,\cite{Hasselba08,Weiss15,Choi07} very few experimental studies providing high-precision measurements of the ring  magnetic moment have been reported.\cite{Jang11} 

Experimentally, the measurement sensitivity for microsample magnetic moments is approaching its limit, as a result of  the notable recent improvement in the force sensitivity in dynamic cantilever magnetometry\cite{Stipe01,Harris99,Chabot03} down to attonewton level.\cite{Bleszyns09,Jang11} In a study of persistent currents in normal metal rings,\cite{Bleszyns09} for example, dynamic cantilever magnetometry, which measures the resonance frequency shift of a cantilever in a magnetic field, exhibited a resolution that was approximately 250-fold superior to SQUID magnetometers\cite{Jariwala01,Bluhm09} for detection of a ring's current. This result finally resolved previous order-of-magnitude discrepancies between experimental and theoretical current values. Such high sensitivity is obtained by applying high external fields. As regards dynamic cantilever magnetometry analysis of the low-field magnetic properties of a sample, however, a significant sensitivity reduction is inevitable. Very recently, this limitation was overcome using a phase-locked approach suggested by Jang et al.\cite{Jang11,Jang11B}  These researchers succeeded in detecting small half-fluxoid-quantum signals in an Sr$_2$RO$_4$ superconductor at low static fields by applying an additional oscillating field, which was phase-locked to the cantilever position, for signal enhancement. The above studies have highlighted the potential sensitivity of dynamic cantilever magnetometry for magnetic-moment detection at both high and low magnetic fields.

In this work, we adopt dynamic cantilever magnetometry for precision measurement of the small magnetic moments of fluxoids in a superconducting microring. However, in order to retain a simple measurement geometry and to reduce the uncertainty factors, we do not employ a phase-locked approach, which requires precise control of the modulation field.\cite{Jang11B} Instead, we enhance the resonance frequency shift by increasing the external field after trapping fluxoids in the ring. For a micron-sized Nb ring, we determine the magnetic moment of a single fluxoid, along with the Meissner susceptibility of the ring, and compare the results with theoretical estimations from the Brandt and Clem method.\cite{Brandt04} For accurate comparison, we prepare a ring sample with a well-defined geometry on an ultrasoft cantilever and utilize a fiber-optic interferometer with subnanometer resolution, with the fiber on a piezo positioner; this setup enables precision vibration measurement at multiple target positions on the cantilever. The latter is necessary for the experimental determination of the cantilever vibration mode characteristics, such as its effective length, which is otherwise theoretically estimated. 

\section{EXPERIMENT}

For the sample-on-cantilever configuration, we batch-fabricate cantilevers with an Nb ring sample. After the cantilever patterns are defined in a low-pressure chemical vapor deposited (LPCVD) silicon-nitride layer on a silicon wafer, a 100-nm-thick Nb ring with nominal inner and outer radii of $a$ = 2 and $b$ = 4 $\mu$m, respectively, is fabricated via lift-off patterning with photolithography. The ring is aligned with the mounting paddle center of each cantilever, as can be seen in Fig.~\ref{fig:schem} (b, right). The cantilever fabrication is then completed, taking care to protect the attached, high-quality Nb film (see Ref.~[\citenum{Choi16}] for more details). The released cantilevers have 367-$\mu$m length, 4-$\mu$m width, and 200-nm thickness, with a mounting paddle at one end. The lateral dimensions and surface quality of the Nb ring are measured and examined using a Tescan Mira scanning electron microscope (SEM). 

The sample-on-cantilever device is placed on a piezoactuator in high vacuum, surrounded by a superconducting solenoid for application of a uniform magnetic field. Its low-temperature vibration amplitude and resonance frequency are measured with a low-noise fiber-optic interferometer using a 1550-nm tunable laser (Agilent 81660B-200) with a high wavelength stability of 1 pm for 24 h and a coherence control feature, which has been demonstrated to have subpicometer resolution at an optical power of 10 $\mu$W and room temperature.\cite{Smith09} For our study, a very low laser power of 13 nW at the fiber end is adopted to avoid optical effects such as photothermal actuation. The fiber, attached to a 3-axis piezo positioner, is located above a target position on the cantilever. The optical interference from the optical-fiber cantilever cavity is detected at a photodiode coupled to a low-noise transimpedance amplifier (Femtoamp DLPCA-200). The cantilever frequency is primarily measured at a temperature of 4.3 K. In the magnetic-field-cooling (FC) process, the cantilever temperature is elevated momentarily using a light-emitting diode to above the superconducting transition temperature, $T_c$, of the Nb ring and then recovered. The entire system is mounted on a double-stage vibration-isolation platform including a 21-ton mass block.

\subsection*{Measurement fundamentals}

\begin{figure*}
\includegraphics*[width=16cm]{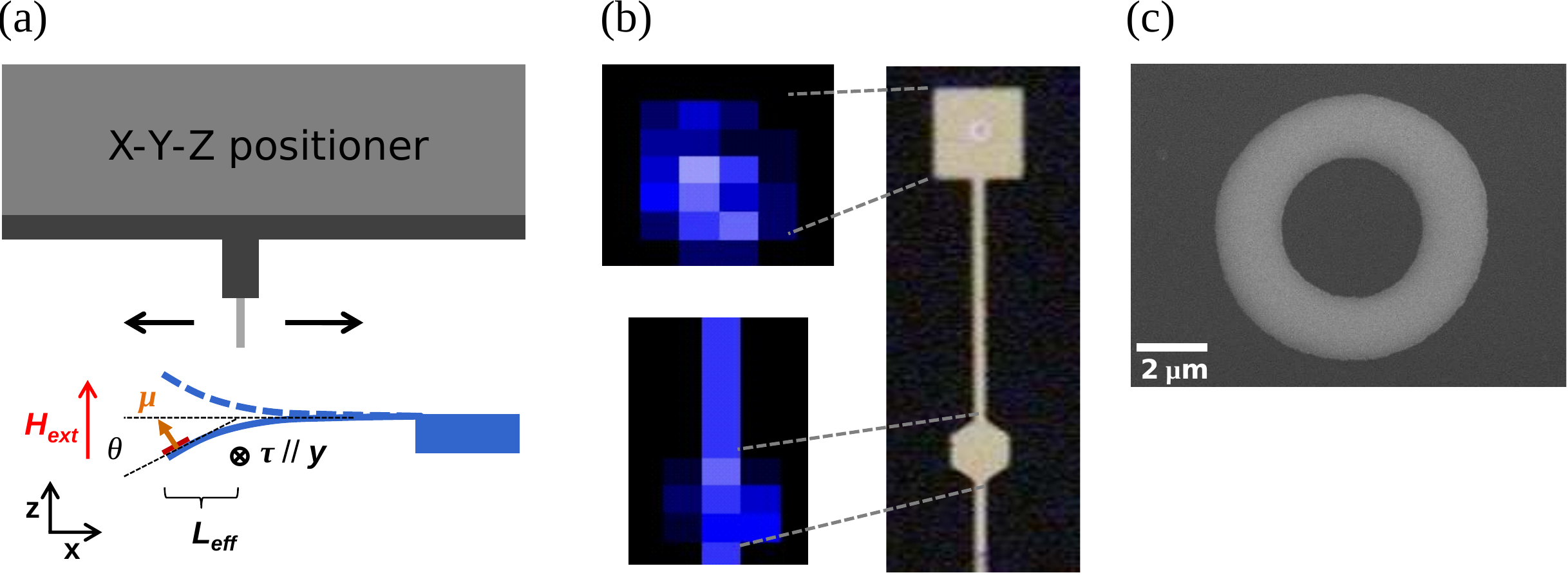}
\caption{(Color online) (a) Schematic of cantilever torque magnetometry. (b) Optical microscope (right) and fiber scan (left) images of a free end part of a silicon nitride cantilever with a paddle (top) and a reflector (middle). (c) Calibrated SEM image of an Nb ring on a paddle. (See Ref.~[\citenum{Choi16}])}
\label{fig:schem}
\end{figure*}

Figure~\ref{fig:schem} (a) shows the key features of our dynamic cantilever magnetometry setup. In an external magnetic field $\bfH_\ext$, the magnetic moment $\bfmu$ of the sample exerts a torque \mbox{\boldmath $\tau = \mu \times \bfH_\ext$} 
on the cantilever. For a two-dimensional sample, we can assume that $\bfmu$ has an out-of-plane component $m$ only. Then, the magnitude of the torque is given as
\begin{equation}
\label{eq:0}
\tau = m H_\ext \sin{\theta}
\end{equation}
and, with $\bfH_\ext \parallel \bfz$, the relative angle $\theta$ of $\bfmu$ and $\bfH_\ext$ is identical to the cantilever surface angle at the sample position with respect to the $\bfx$ direction. 
% Thus, the torque is a function of the cantilever vibration.

The shift of the resonance frequency $f$ due to the magnetic torque\cite{Stipe01,Jang11B} is expressed as
\begin{equation}
\label{eq:1}
\Delta f = \frac{f_0}{2 k_0 L_\eff^2} m H_\ext =  \frac{f_0}{2 k_0 L_\eff^2} \left( \chi H_\ext+n m_\FQ \right) H_\ext  ,
\end{equation}
where $k_0$ and $f_0$ are the spring constant and intrinsic resonance frequency of the cantilever, respectively, and $L_\eff$ is the cantilever effective length.
%$\omega_0$  angular resonance frequency $2\pi f_0$ (where $f_0$ is the intrinsic resonance frequency)
In the case of our superconducting ring, $m$ has two contributions, from the diamagnetic response due to the Meissner current and from the $n$ magnetic fluxoids in the ring hole. Here, $\chi$ is the Meissner susceptibility, and each fluxoid quantum has the same magnetic moment, $m_\FQ$.

In our work, we adopt a cantilever, shown in Fig.~\ref{fig:schem} (b), with $f_0$ = 1221.9 Hz and $k_0 = 4.5\times10^{-5}$ N/m in the fundamental vibrational mode. For $L_\eff$, the theoretical value of $L/1.38$ for a rectangular Euler-Bernoulli beam of length $L$ is frequently used.\cite{Sidles95} However, the $L_\eff$ of our device was experimentally determined to be $L/1.48$ or 248 $\mu$m, by measuring the shape of the first vibration mode with a fiber on the piezo positioner. The minimum detectable frequency shift and magnetic moment, $\Delta f_\mmin$ and $m_\mmin$, respectively, of our cantilever were estimated to be 1.1 mHz and 1.2 fAm$^2$, respectively, for a 1-Hz detection bandwidth with $H_\ext$ = 40 Oe. The characterization of the cantilever mechanical properties is described in more detail in Appendix A.

\section{RESULTS AND DISCUSSION}

\begin{figure}
\includegraphics*[width=\columnwidth]{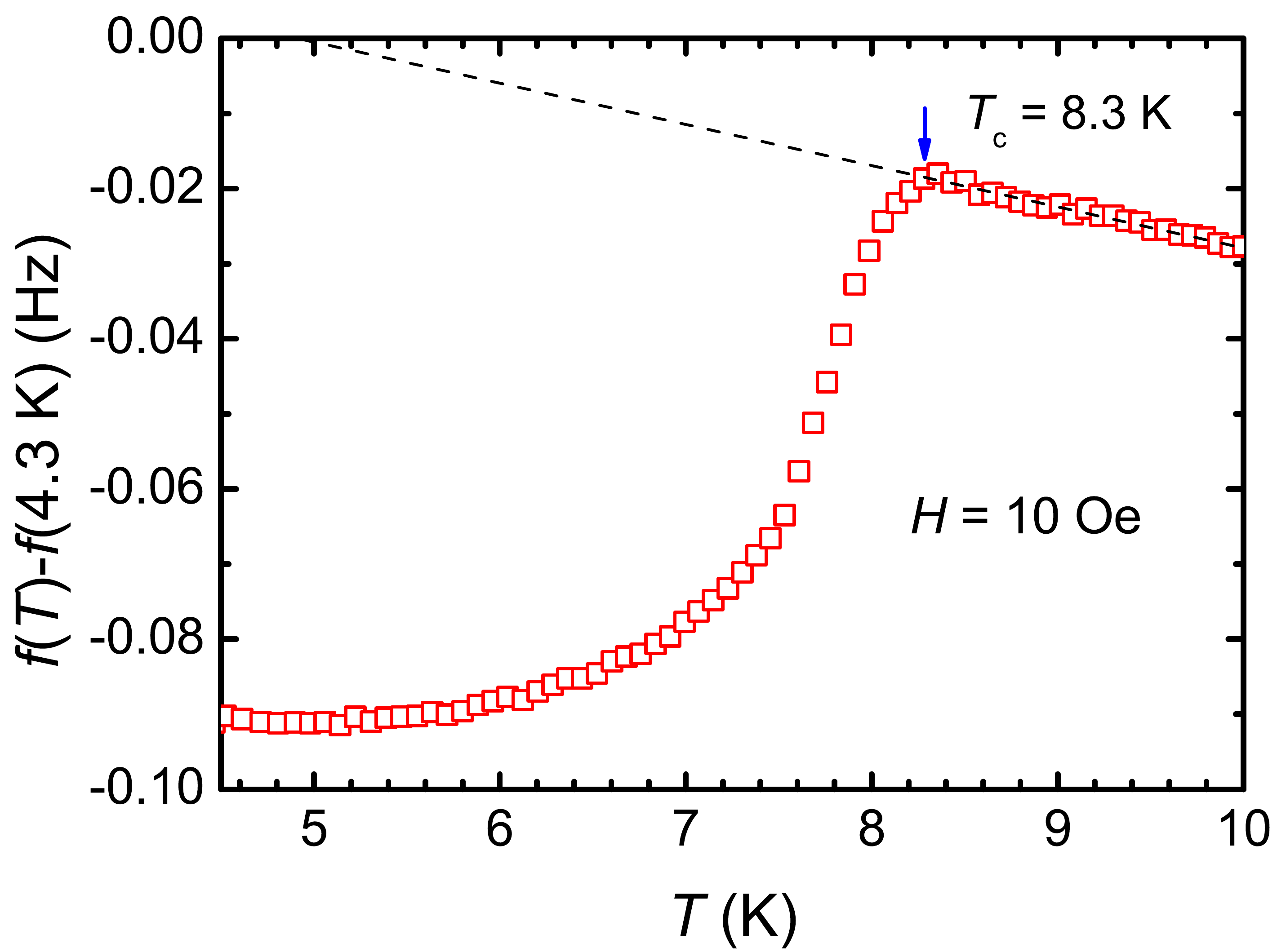}
\caption{(Color online) Temperature dependence of the resonance frequency $f$ of the Nb-ring-mounted cantilever in a magnetic field of 10 Oe, applied along the $z$-direction after zero-field cooling. The arrow indicates the diamagnetic onset temperature of superconductivity. The dashed line is a linear fit to the data above $T$ = 8.4 K.}
\label{fig:fvsT}
\end{figure}

To observe the superconducting transition, the $f$ of the cantilever was monitored with increasing temperature in a magnetic field of 10 Oe, applied perpendicularly to the mounted Nb ring after zero-field cooling to $T =$ 4.5 K. The $f$ temperature dependence exhibits a typical feature of a diamagnetic superconducting transition, with an onset temperature of $T_c =$ 8.3 K, with the exception that a slope is apparent across the entire displayed temperature range (Fig.~\ref{fig:fvsT}). This feature indicates that the superconducting ring is in the Meissner state at temperatures lower than $T_c$. The $T_c$ value agrees well with the superconducting transition temperature obtained for a resistive measurement of a strip Nb sample from the same batch (data not shown here).

The  $f_0$ in the absence of $\tau$ is represented by a dashed line in Fig.~\ref{fig:fvsT}, having a slope of -5.5 mHz/K; this slope is obtained from a fit of the data in the normal state. The possible origins of the negative slope are the temperature dependence of the cantilever dimensions, cantilever surface stress, and so on; further discussion of this topic is presented in Appendix B. 

\begin{figure*}
\includegraphics*[width=17cm]{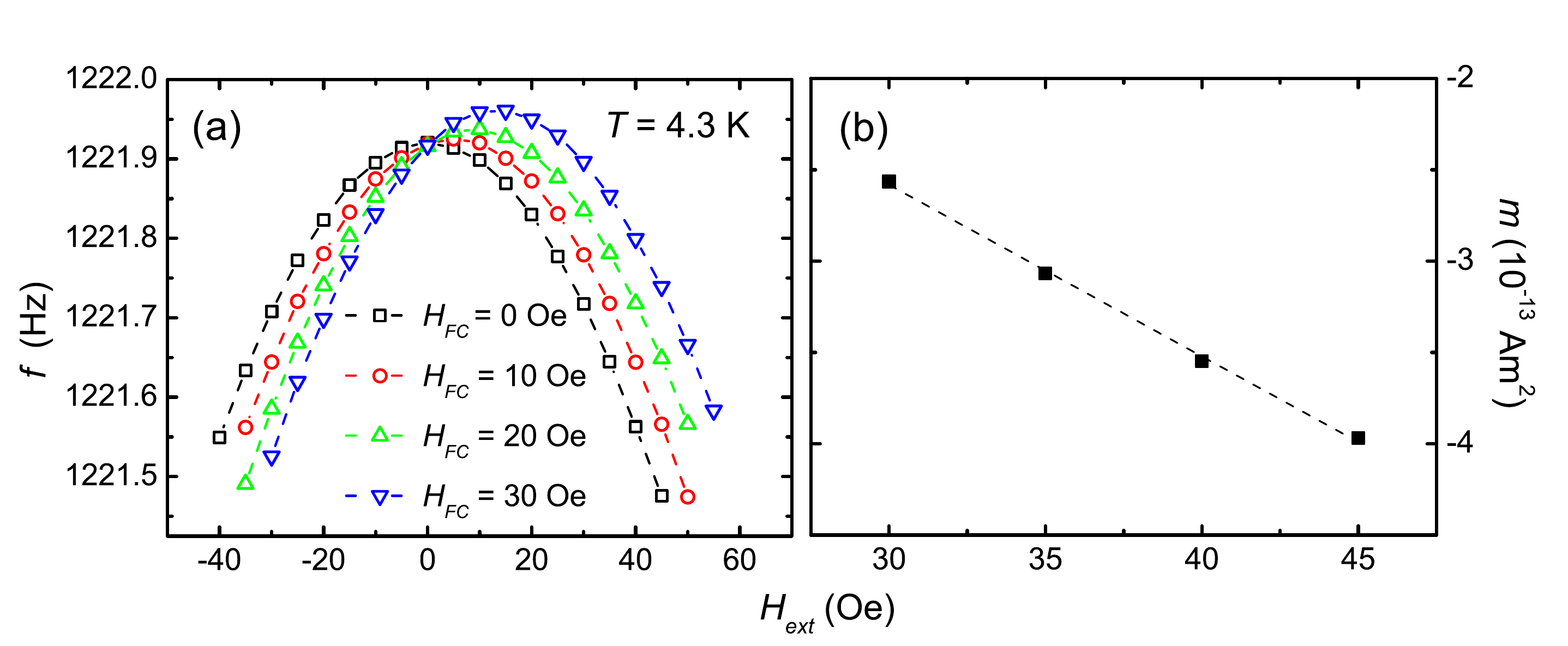}
\caption{(Color online) (a) Resonance frequency $f$ of the cantilever versus the external magnetic field, $H_\ext$, for various values of $H_\FC$, the magnetic field used in field-cooling. (b) Diamagnetic response for increasing $H_\ext$, calculated from the $f$ shift for $H_\FC = 0$ Oe. The dashed line is a linear fit.}
\label{fig:Hdep}
\end{figure*}

In the Meissner state of the Nb ring, the $f$ response to sweeping $H_\ext$ follows Eq.~(\ref{eq:1}), resulting in the parabolic curve shown in Fig.~\ref{fig:Hdep} (a). The parabolic dependence is valid in the  $\left| H_\ext \right| \leq \sim$ 60 Oe range, whereas it breaks down beyond this range as a result of magnetic vortex penetration into the annular area, i.e., a mixed state of Nb. Such a small critical field value is attributed to the high demagnetization effect due to the quasi-2D sample geometry.\cite{Doria08} As we increase the FC magnetic field, $H_\FC$, used in cooling the Nb ring from above $T_c$, more magnetic fluxoids are contained within the ring hole. Accordingly, the curve is shifted to higher $f$ and $H_\ext$.
%% A crossing point of curves represents the state at which the net external magnetic field, not necessarily $H_\ext$, is zero. Therefore, the  background field $H_\BG$, e.g., a remnant field trapped in the superconducting magnet of our setup, can be deduced from this point and, in principle, used for field correction. By enlarging Fig.~\ref{fig:Hdep} (a) (not shown), $H_\ext$ for the crossing point (or $-H_\BG$) is found to be 0.5 Oe, with a large error of 0.3 Oe. Because $H_\BG$ is small and comparable to its error, this field is not considered in the data plots shown in this work. 

Parabolic fits to the data shown in Fig.~\ref{fig:Hdep} (a) can provide $\chi$ and $n m_\FQ$; however, we obtained the $n m_\FQ$ values from separate measurements, which proved to be more accurate and efficient. We deduced  $\chi H_\ext$ by dividing the $H_\FC$ = 0 data by $H_\ext$, which exhibits a linear dependence, as depicted in Fig.~\ref{fig:Hdep} (b). Note that data at low magnetic fields were not employed, because of their low accuracy. The linear fit yields $\chi  = -102 \pm 10$ pAm$^2$/T.

\begin{figure}
\includegraphics*[width=\columnwidth]{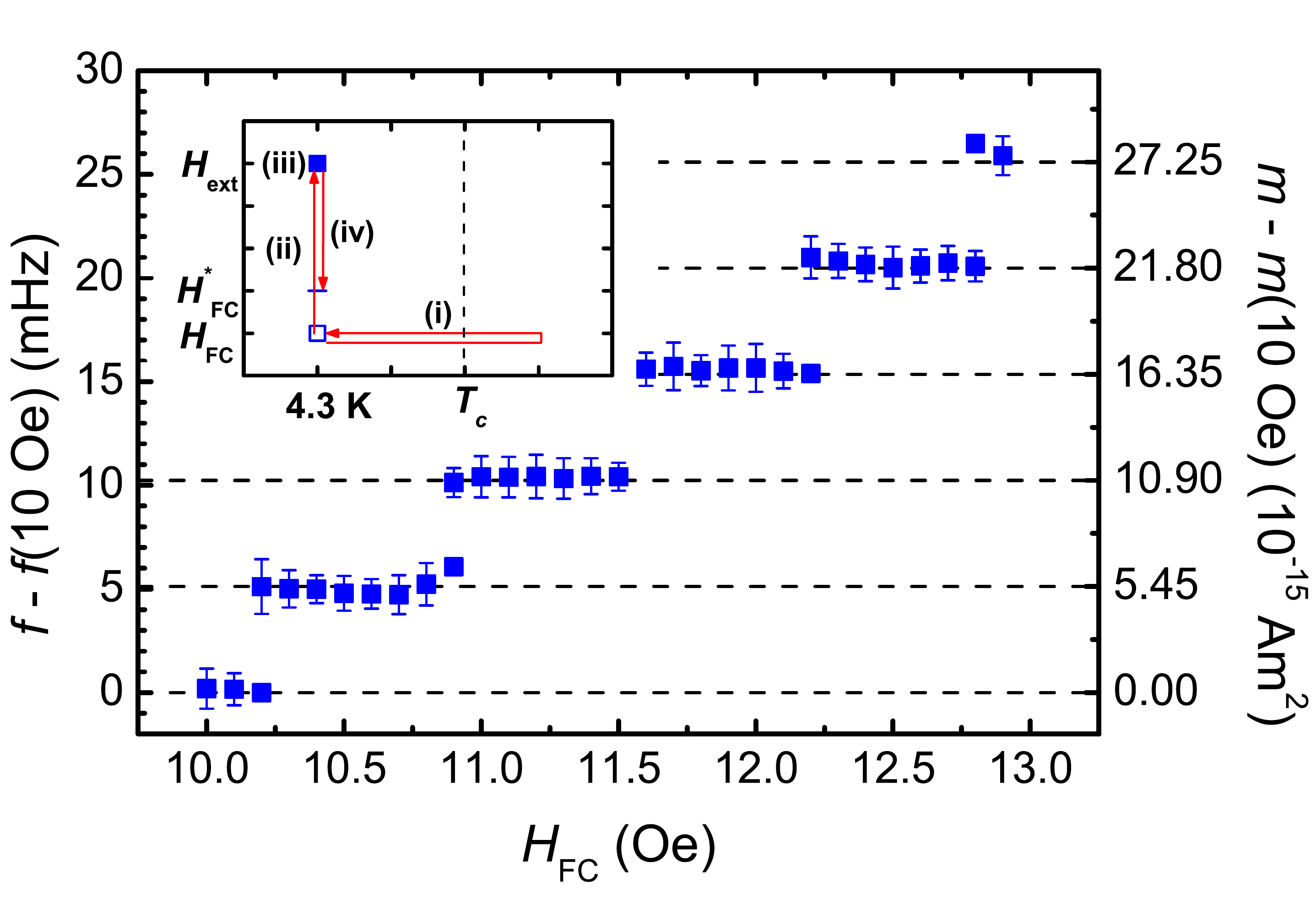}
\caption{(Color online) Resonance frequency shift (left axis) and magnetic moment (right axis) vs. $H_\FC$. Each data point was obtained from five repetitions of a temperature-field cycle and a single measurement with an averaging time of 40 s, described in the inset: (i) warm-up and field cooling at $H_\FC$; (ii) increasing field up to $H_\ext$; (iii) $f$-shift measurement; and (iv) decreasing field to the next value, $H_\FC^{*}$. The error bars indicate standard deviations, and the dashed lines are guides for the eye.}
\label{fig:fstep}
\end{figure}

To observe individual magnetic fluxoids at $T =$ 4.3 K, we varied $H_\FC$ from 10 to 13 Oe with a smaller step of 0.1 Oe. To enhance the $\Delta f$ signal for $n m_\FQ$ for low $H_\FC$, we increased the magnetic field from $H_\FC$ to a larger and fixed value, i.e.,$H_\ext$ = 40 Oe, before measuring $f$. This procedure is depicted in the inset of Fig.~\ref{fig:fstep}. In this manner, we could obtain $n m_\FQ$ for small $n$, because $n m_\FQ$ is independent of the magnetic field, but its contribution to $\Delta f$ is proportional to $H_\ext$, as shown in Eq.~(\ref{eq:1}). Note that the contributions of $\chi H$ for various $H_\FC$ are identical and can be universally eliminated because $H_\ext$ is fixed. Figure~\ref{fig:fstep} clearly shows that $f$ has a stepwise feature with varying $H_\FC$. The single step width, $\Delta H$, was estimated to be 0.65 $\pm$ 0.03 Oe from the total width of the four steps fully shown in Fig.~\ref{fig:fstep}. Taking the errors in $H_\FC$ and $\Delta H$ into consideration, the $n$ corresponding to $H_\FC$ = 10 Oe may range from 14 to 16.

For FC with a $H_\FC$ corresponding to the center of each step plateau, no net current circulates the ring, even with $n$ fluxoids and the response to the external field.  The effective area of the zero-current contour is given by $A_\eff = \Phi_0/H_\mathrm{a}$,\cite{Brandt04} where $H_\mathrm{a}$ is the field increment necessary to induce a transition from the $n$ to $n + 1$ state.  Because $H_\mathrm{a} = \Delta H$, we can estimate $A_\eff$ to be 32 $\mu$m$^2$, which indicates flux focusing where $A_\eff$ is larger than the actual hole area, $\pi a^2 =$ 13 $\mu$m$^2$. This estimate agrees roughly with the $A_\eff =$ 25 $\mu$m$^2$ result calculated using the Brandt and Clem theoretical prediction.\cite{Brandt04}

Within $\Delta H$, $f$ is virtually constant to within 1 mHz for changing $H_\FC$, which implies that the number of fluxoids is fixed and their contribution to $m$ is constant. As $H_\FC$ is raised beyond $\Delta H$, an additional fluxoid is introduced to the ring hole, resulting in a discrete shift of $f$ or $m_\FQ$ as shown in Fig.~\ref{fig:fstep}. As the $m_\FQ$ of each fluxoid are intrinsically expected to be equivalent, this value can be determined from the average of the five steps of $m$ or $f$, which are $5.8\pm0.6$ fAm$^2$ or $5.2\pm0.2$ mHz, respectively. Note that the uncertainty in the cantilever spring constant makes a dominant contribution to the estimated error in $m_\FQ$. Near the step edges, $f$ is observed at both fluxoid quantum numbers, $n$ and $n + 1$, because the kinetic energies of the right- and left-circulating supercurrent states, respectively, are degenerate for Nb-ring cooling at corresponding magnetic fields. 

\begin{figure*}
\includegraphics*[width=17cm]{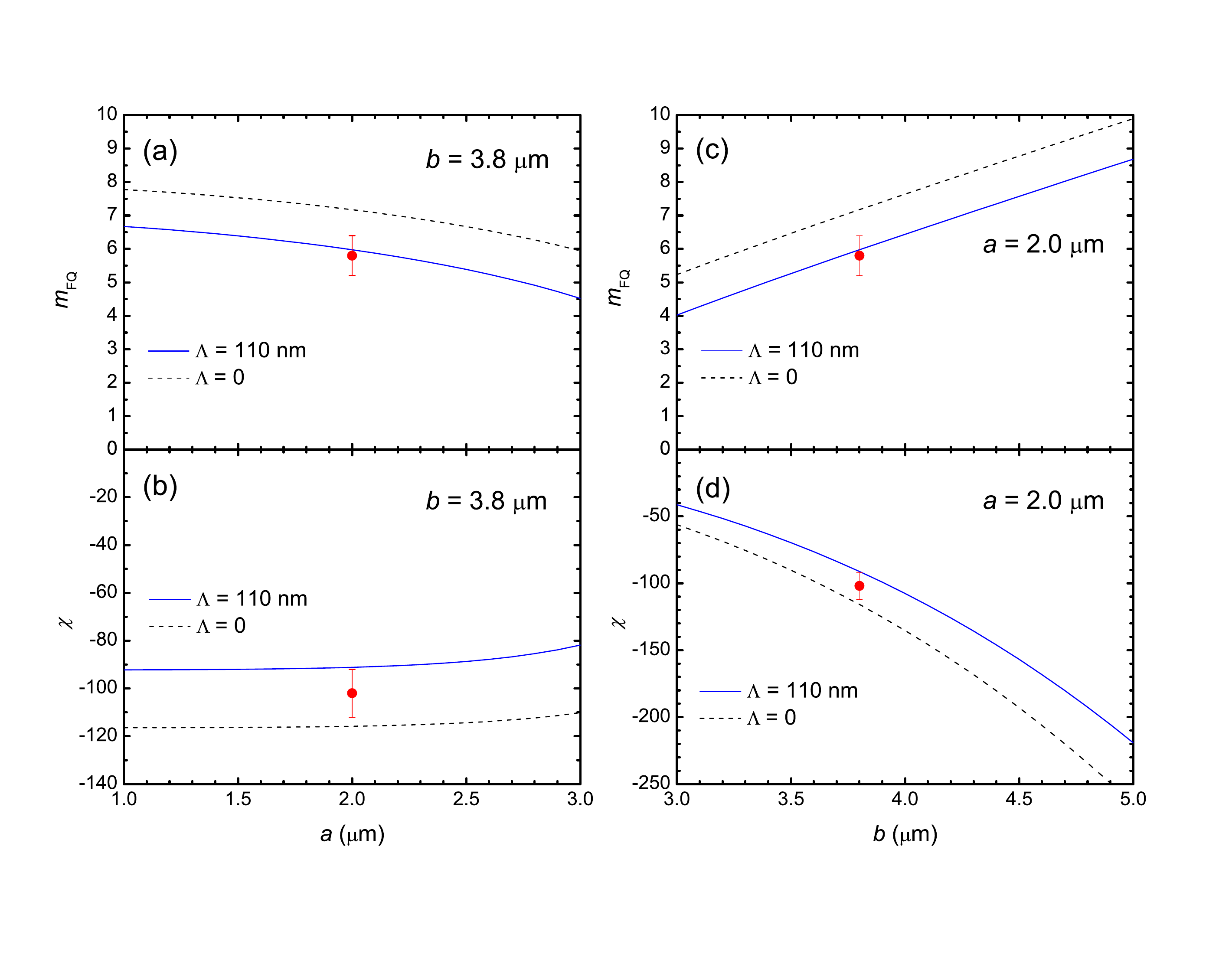}
\caption{(Color online) Theoretical estimates of $m_\FQ$ and $\chi$ for $\Lambda = 0$ and 110 nm for a superconducting ring with varying inner radius $a$ (a \textendash \:b) and outer radius $b$ (c \textendash \: d). The circles represent the experimental values with error bars indicating 10\% accuracy.}
\label{fig:theory}
\end{figure*}

Figure~\ref{fig:theory} depicts the theoretical values of the magnetic moments due to a single fluxoid and a diamagnetic response, which were estimated numerically for various ring radii utilizing the Brandt and Clem model.\cite{Choi07,Brandt04} As shown in the figure, the magnitudes of $m_\FQ$ and $\chi$ decrease slowly with increasing $a$, but increase with higher dependence with increasing $b$. The experimentally obtained values are also plotted at the dimensions of our Nb ring on the cantilever; the dimensions are measured from the calibrated SEM, yielding $a$ and $b$ of $2.0 \pm 0.1$ $\mu$m and $3.8 \pm 0.1$ $\mu$m, respectively. For $\Lambda =$ 110 nm at 4 K,\cite{Kim12} where $\Lambda =\lambda^2/d$ is the thin-film penetration depth and $d$ is the film thickness,  $m_\FQ$ and $\chi$ are calculated to be $5.98 \pm 0.10$ fAm$^2$ and $-91.2 \pm 0.3$ pAm$^2$/T ($5.98 \pm 0.24$ fAm$^2$ and $-91.2 \pm 7.8$ pAm$^2$/T), respectively, if the uncertainty of $a$ ($b$) is considered. These values are in quite good agreement with the experimental results, considering the accuracy of the spring constant determination.

The effect of the $\Lambda$ uncertainty is negligible, as $m_\FQ$ and $\chi$ are estimated to vary by only 1.4\% and -1.7\%, respectively, for a $\Lambda$ difference of 10\%. However, it is notable that the assumption of negligible $\Lambda$ in the theoretical estimation yields $m_\FQ$ = 7.17 fAm$^2$ and $\chi$ = -116 pAm$^2$/T, which are considerable overestimations in comparison with the experimental values. This finding implies that consideration of the finite penetration depth, as in the Brandt and Clem model, is crucial for appropriate description of micron-sized superconducting rings, and that it remains valid when the ring radii uncertainty is considered, as can be seen in Fig.~\ref{fig:theory}.

\section{CONCLUSIONS}
Using high-resolution cantilever magnetometry capable of fiber scanning, we precisely measured the magnetic moment of a well-defined superconducting Nb thin-film ring with inner and outer radii of 2.0 $\mu$m and 3.8 $\mu$m, respectively, on an ultrasoft cantilever at $T$ = 4.3 K. The experimental results, a diamagnetic response of -102 pAm$^2$/T and a single fluxoid magnetic moment of 5.8 fAm$^2$, agree well with the theoretical model prediction, providing a reliable technical and theoretical base for superconducting microring research and applications in the future. 

%%%%%%%%%%%%%%%%%%%%%%%

%%%%%%%%%%%%%%%%%%%%%%%

\begin{acknowledgments}
The authors are grateful to D. H. Lee and B. H. Park for SEM imaging. This work was supported by the Korea Research Institute of Standards and Science under the ''Establishment of National Physical Measurement Standards and Improvements of Calibration/Measurement Capability'' project, Grant Nos. 16011003 and 16011012. M.-S. C. was supported by the National Research Foundation of Korea (Grant No. 2015-003689).
\end{acknowledgments}

\appendix
\renewcommand\thefigure{\thesection.\arabic{figure}}  
%\chapter{Appendix}
\section{}
\setcounter{figure}{0}

In dynamic cantilever magnetometry, the resonance-frequency shift $\Delta f$ can be derived by calculating the magnetic torque oscillation dependent on the cantilever vibrations.  The cantilever vibration is a solution of the equation of motion for beam vibration,\cite{Rao95} which is generally expressed as
\begin{equation}
\label{eq:A1}
u(x,t) = \sum_{n=1}^{\infty} u_n (x) q_n (t) ,
\end{equation}
where $u_n(x)$ is the $n$th resonance mode shape and $q_n (t)$ is the generalized coordinate in the $n$th mode. If we drive the cantilever at one of the resonance frequencies, for example, the first mode, the problem is reduced to solving a one-dimensional forced equation of motion for $q_1 (t)$. The cantilever is subject to an effective force of $\tau/L_\eff$, where $L_\eff$ is the cantilever effective length, defined as $u_1 (x)/\tan{\theta} = u_1 (x)/(du_1 (x)/dx)$.\cite{Sidles95} The effective force can then be deduced as 
\begin{equation}
\label{eq:A2}
\frac{\tau}{ L_\eff} = \frac{1}{ L_\eff^2} m H_\ext  u_1 (x) q_1 (t) 
\end{equation}
from Eq.~(\ref{eq:0}),
with an approximation of $\sin{\theta} \cong  \frac{du_1 (x)}{dx} q_1 (t) = \frac{u_1 (x)}{L_\eff} q_1 (t)$ 
for small deflections. Hence, the Fourier transform of the forced vibration equation for $q_1(t)$ can be expressed as\cite{Jang11B} 
\begin{equation}
\label{eq:A3}
(-\omega^2 -i \gamma \omega + \omega_0^2) \tilde{q} (\omega) = \frac{\omega_0^2}{k_0 L_\eff^2} m H_\ext \tilde{q} (\omega) .
\end{equation}
Here, $\omega_0$ is the angular resonance frequency $2\pi f_0$. The solution of Eq.~(\ref{eq:A3}) gives $\Delta f$ as expressed in Eq.~(\ref{eq:1}).\cite{Stipe01,Jang11B}

As illuminating the cantilever free end, even at small laser power, may cause local heating of the sample, $\Delta f$ measurement for magnetometry is conducted with the fiber pointing at the center of a 20 $\mu$m-width reflector, shown in Fig.~\ref{fig:schem} (b), ~100 $\mu$m from the paddle on which an Nb ring is mounted. To align the fiber to the reflector center or another point of interest, we first obtain a quick map of the cantilever, as shown in Fig. 1(b), by scanning the cantilever plane and obtaining the laser interference amplitude at each point; this is achieved by sweeping the fiber-cantilever inter-distance. Then, for fine adjustment, we repeatedly obtain line profiles of the interference amplitude, in directions both parallel and perpendicular to the cantilever, to find the target position with $\sim1$ $\mu$m resolution.

To determine precise values for $k_0$ and $L_\eff$ in Eq.~(\ref{eq:1}), we require a fundamental mode shape; therefore, we obtain position-dependent vibrational noise spectra along the cantilever. These spectra provide $\left< u^2 (x,t) \right>$ from Eq.~(\ref{eq:A1}), which falls on the mode shape predicted by the finite element method for the cantilever employed in this work. From the ratio of $\left<u^2 (x,t) \right>$ at the sample position, $\left< u_\bfS^2 \right>$, against that at the reflector center, $\left< u_\bfR^2 \right>$, we determine the spring constant conversion factor, $\left< u_\bfS^2 \right> / \left< u_\bfR^2 \right>$, to be 2.85, and from the slope at the sample position, we determine $L_\eff$ to be 248 $\mu$m.

\begin{figure}
\includegraphics*[width=\columnwidth]{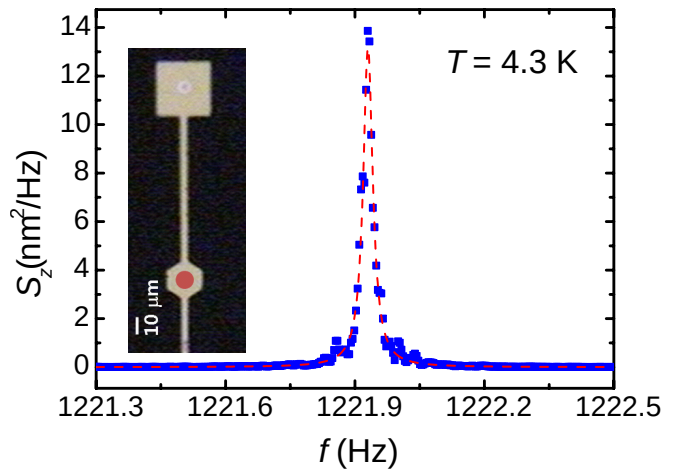}
\caption{(Color online) Thermal vibration noise spectrum at $T$ = 4.3 K measured at the reflector position (closed circle in the inset).}
\label{fig:spect}
\end{figure}

Figure~\ref{fig:spect} shows the fundamental thermal vibration noise spectrum at $T$ = 4.3 K, obtained with a span of 3.125 Hz and averaging over 15 results, which provides $\left< u_\bfR^2 \right>$ as well as $f_0 =$ 1221.9 Hz and the quality factor $Q =$ 43000. Using the equipartition theorem along with $\left< u_\bfS^2 \right> / \left< u_\bfR^2 \right>$, the mechanical impedance to the force at the sample position is evaluated to be $k_0 = 4.5 \times 10^{-5}$ N/m, with an accuracy conservatively claimed to be 10\%.\cite{Matei06} The minimum detectable shift of the cantilever frequency is given by $\Delta f_\mmin = f_0 F_\mmin/ \sqrt{2} k_0 x_\pk$.\cite{Stipe01} Here, $F_\mmin$ is the smallest detectable force signal, given by $F_\mmin =  \sqrt{2k_0 k_\mathrm{B} T B / \pi f_0 Q}$, where $k_\mathrm{B}$ is the Boltzmann constant, $x_\pk$ is the peak displacement of the oscillating cantilever, and $B$ is the detection bandwidth. The thermally limited detectable magnetic moment $m_\mmin$ can be expressed as 
%$2 \left( \frac{\Delta f_\mmin}{f_0} \right) \frac{k_0 L_\eff^2}{H_\ext}$ , 
$2 \Delta f_\mmin k_0 L_\eff^2/ f_0 H_\ext $,
employing Eq.~(\ref{eq:1}). Using the cantilever parameters given above, the corresponding $\Delta f_\mmin$ and $m_\mmin$ are 1.1 mHz and 1.2 fAm$^2$ for a 1-Hz bandwidth with $x_\pk$ = 100 nm and $H_\ext$ = 40 Oe.

\section{}
The negative slope of $f_0 (T)$ in Fig.~\ref{fig:fvsT} may originate from the temperature dependence of the Young's modulus, dimensions, surface stress, and so on, of the silicon nitride cantilever. The spring constant of a simple beam is given by\cite{Sidles95} $k_0 = 1.030 E w t^3 / l^3$, where $E$ is the Young's modulus of the material and $w$, $t$, and $l$ are the beam width, thickness, and length, respectively. With $2 \pi f_0 = \sqrt{k_0/m_\eff}$, where $m_\eff$ is the beam effective mass, the temperature derivative of $f_0 (T)$ can be expressed as 
\begin{equation}
\label{eq:B1}
\frac{1}{f_0} \frac{df_0}{dT} = \frac{1}{2 k_0} \frac{dk_0}{dT} = \frac{1}{2} \left( \frac{1}{E} \frac{dE}{dT} + \frac{1}{w} \frac{dw}{dT} \right) ,
\end{equation}
where we assume an isotropic thermal contraction for $w$, $t$, and $l$. 

The effect of the intrinsic Young's modulus can be ignored because, in general, its temperature dependence is virtually zero at low temperatures.  If we adopt the Wachtman semi-empirical formula for Young's modulus,\cite{Bruls01} $E(T)=E_0 - B T \exp (-T_0/T)$, its temperature derivative is given by $dE/dT= -B(1+ T_0/T) \exp (-T_0/T)$. For the reported parameters for silicon nitride,\cite{Bruls01} $E_0 =$ 320 Gpa, $B$ = 0.0151 GPa/K, and $T_0$ = 445 K, $(1/E) dE/dT$ is estimated to be as small as $-1 \times 10^{-24}$ K$^{-1}$ at $T$ = 9 K. 

Excluding the intrinsic Young's modulus, we may speculate that the temperature dependence of the cantilever dimensions yields the $f_0 (T)$ slope both indirectly and directly, via the first and second terms on the right-hand side of Eq.~(\ref{eq:B1}), respectively. One possible indirect effect is via surface stress in a thin cantilever. Because of the strain-dependent surface stress, the effective Young's modulus $E_\eff$ of a silicon nitride cantilever has been reported to have a thickness dependence.\cite{Gavan09} That is, $E_\eff$ decreases strongly for decreasing thickness below our cantilever thickness of 200 nm.

Considering the thickness dependence and the signs in Eq.~(\ref{eq:B1}), the thermal contraction of the cantilever dimensions for increasing $T$ is consistent with the negative slope of $f_0 (T)$, if other factors are ignored.  The lower bound of the thermal expansion coefficient $\alpha$, which is defined as $\alpha = (1/w) dw/dT$, can be estimated from Eq.~(\ref{eq:B1}) with the assumption of $dE/dT = 0$, yielding $\alpha_\mathrm{lower} = (2/ f_0) d f_0 / dT = -9 \times 10^{-6}$ K$^{-1}$. More systematic studies are necessary in the future to determine an accurate value of $\alpha$ for silicon nitride at low temperatures.

%%%%%%%%%%%%%%%%%%%%%%%


\end{document}